\documentclass[a4paper,fleqn,usenatbib]{mnras}

\usepackage[T1]{fontenc}
\usepackage{ae,aecompl}

\usepackage{graphicx}	
\usepackage{amsmath}	
\usepackage{amssymb}	
\usepackage {threeparttable} 

\newcommand{\msunyreq}{\mathrm{M}_{\sun}\,\mathrm{yr}^{-1}}

\title[Perseus cluster X-ray minicoronae]{The X-ray Coronae of two massive galaxies in the core of the Perseus cluster}

\author[N. Arakawa et al.]{
N. Arakawa,$^{1, 2}$\thanks{E-mail: na489@cam.ac.uk}
A. C. Fabian$^{1}$
and S. A. Walker$^{3}$
\\
$^{1}$Institute of Astronomy, University of Cambridge, Madingley Road, Cambridge CB3 0HA, UK\\
$^{2}$Kavli Institute for Cosmology, University of Cambridge, Madingley Road, Cambridge CB3 0HA, UK\\
$^{3}$Astrophysics Science Division, X-ray Astrophysics Laboratory, Code 662, NASA Goddard Space Flight Center, Greenbelt, MD 20771, USA\\
}

\date{Accepted 2019 June 25}

\pubyear{2019}

\begin{document}
\label{firstpage}
\pagerange{\pageref{firstpage}--\pageref{lastpage}}
\maketitle

\begin{abstract}
We study the X-ray properties of two elliptical galaxies, NGC 1270 and NGC 1272, in the core of the Perseus cluster with deep {\it Chandra} observations. Both galaxies have central supermassive black holes, the mass of which is $6.0 \times 10^{9}\ \mathrm{M}_{\sun}$ and $2.0 \times 10^{9}\ \mathrm{M}_{\sun}$ respectively. Our aim is to examine relatively cool soft X-ray emitting gas within the central region of these massive early-type galaxies. Such gas, referred to as a Minicorona in previous studies is common in the core of large elliptical cluster galaxies. It has not been completely stripped or evaporated by the surrounding hot intracluster medium and nor fully accreted onto the central black hole. With thermal emission from the minicorona dominating over any power-law radiation components, we find that both NGC 1270 and NGC 1272 encompass minicoronae, the temperature and radius of which are $0.99$ keV and $0.63$ keV; $1.4$ kpc and $1.2$ kpc respectively. For NGC 1272, the thermal coronal component dominates the core emission by a factor of over 10. We show that the depletion time scale of minicoronal gas via viscous stripping is shorter by a factor of $100$ than the replenishment time scale due to stellar mass loss. Magnetic fields are presumably responsible for suppression of the transport processes. Finally, we show that both objects have to meet a balance between cooling and heating as well as that among mass replenishment, stripping and accretion.
\end{abstract}





\section{Introduction}

It is expected that the intracluster medium (ICM) interacts with the interstellar medium (ISM) in member galaxies of clusters. Therefore, it might be expected that the ISM in elliptical galaxies in the core of a cluster would be evaporated or stripped away by the hot ICM surrounding the galactic components as each galaxy moves in a high-velocity orbit in the relatively static ICM of the cluster core. However, against this expectation, X-ray observational studies of nearby clusters have shown that many elliptical galaxies still contain a compact central region of relatively cooler and denser gas. This compact gas typically has a temperature of about 1 keV and radius of up to a few kiloparsecs centered on their massive black holes \citep{2005ApJ...619..169S,	2002ApJ...578..833Y,	2001ApJ...555L..87V}. Most of the rest of the ISM could have been stripped, but that in the deepest part of the potential appears to survive even in the harsh ICM environment \citep[][thereafter this is referred to S07 in this paper]{2007ApJ...657..197S}. In this paper, we refer to this relatively cool compact gas region as a minicorona in order to distinguish them from the typical outer ISM, since that is likely to have been removed. 

Because of their $\sim1$ keV temperature, minicoronae are observed as extended regions of mainly soft X-ray emission with some conspicuous thermal emission line features. This makes them spectrally separate from unresolved low mass X-ray binaries (LMXBs) in the rest of the galaxy.

It might be expected that the minicoronal gas would fuel the central black holes of the galaxies to create accretion-powered luminosity. Nevertheless, they often appear to be under-luminous in comparison with such accretion-powered luminosities \citep{2007MNRAS.382..895S}. This characteristic is generally in accordance with those of galaxies containing supermassive black holes at their cores \citep{1988Natur.333..829F}. It typically shows that the accretion is likely to be both inefficient as a process and in producing radiation \citep[e.g.][]{2005ApJ...624..155P, 	2006ApJ...640..126S,	2011ApJ...736L..23W}.

These facts bring several significant questions: where did minicoronae originate, and why can the gas stay in the cores of the galaxies? For the former question, one obvious explanation is stellar mass loss. For the latter one, the reason why they have not been stripped, evaporated or accreted onto the central supermassive blackhole is not yet well understood. Furthermore, the cooling and heating balance of minicorona remains unclear.

 Here, we study the incidence of minicoronae and their X-ray and physical properties in two elliptical galaxies in the core of the Perseus cluster: NGC 1270 and NGC 1272, using deep {\it Chandra} observations. They are chosen, following earlier work on NGC 1277 \citep{2013MNRAS.431L..38F}, since they are inferred to have massive central black holes, and their location in the Perseus cluster means that deep archival data are available. Previously, four minicoronae were identified in the Perseus cluster by \citet{2007MNRAS.382..895S} by the presence of thermal spectral components using a combined $890$ ks image of the cluster core consisting of $13$ ACIS-S observations, while these observations, combined with ACIS-I and ACIS-S images, were also analyzed by S07 resulting in the identification of nine minicoronae in the Perseus cluster. NGC 1272 was not included in either previous study. NGC 1270 is noted by S07 with similar luminosity and temperature as shown here.
 
The results of our analysis are used to show the existence of minicoronae in the cores of NGC 1270 and NGC 1272 as well as to describe their physical properties. This leads to the discussion where replenishment time is compared with depletion time scale of the minicorona and whether their observed AGN activities and radiation efficiencies are high or low compared to their theoretical accreting luminosity.

This paper is structured as follows: the targeted objects in the {\it Chandra} X-ray observation images, spectral analysis and surface brightness profiles are described in Section~\ref{sec:data}. Minicorona identification, physical properties of minicorona, replenishment and depletion processes, and accretion-powered bolometric luminosity are shown in Section~\ref{sec:results}. Finally, the results are discussed in Section~\ref{sec:discussion}.

It is assumed that $H_0=70$ km s$^{-1}$ Mpc$^{-1}$, $\Omega_{\mathrm{m}} = 0.3$ and $\Omega_\Lambda =0.7$. The redshifts of these two galaxies in the Perseus cluster are assumed to be equivalent to the redshift of the central galaxy NGC 1275, which is 0.0179. All errors unless stated otherwise are at the $1\,\sigma$ level.

\begin{figure}
	\includegraphics[width=\columnwidth]{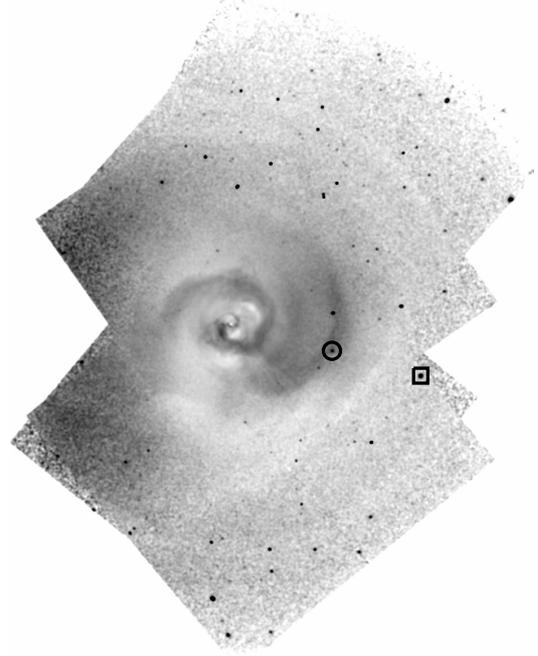}
    \caption{$0.5-7.0$ keV consolidated and exposure-corrected image of {\it Chandra} observations of the Perseus cluster. NGC 1272 is marked with a circle while NGC 1270 is indicated with a square.}
    \label{fig:chandra}
\end{figure}

\section{Data analysis}
\label{sec:data}
\subsection{X-ray images}
The {\it Chandra} images of the Perseus cluster analyzed in this paper are described by \citet{2011MNRAS.418.2154F}. Fig.~\ref{fig:chandra} shows the consolidated image of the Perseus cluster in $0.5-7.0$ keV labelling the targeted two elliptical galaxies: NGC 1270 and NGC 1272. The 8 ACIS-I $17.9-112.2$ ks images were taken in 2009 where each of them covered an area of $16 \times 16$ arcmin. As the result, the combined image covers an area of roughly 36 arcmin by 30 arcmin centred on the core of the Perseus cluster. These eight {\it Chandra} image identification numbers are 11713, 11714, 11715, 11716, 12025, 12033, 12036 and 12037. A catalogue of the Perseus cluster galaxies given by \citet{1999AampAS..139..141B} was used to identify these two X-ray sources. The coordinates of NGC 1270 and NGC 1272 are RA (J2000): $49.7419$; DEC (J2000): $+41.4702$ and RA (J2000): $49.8385$; DEC (J2000): $+41.4908$, respectively. The effective exposure time of the observation is $1.98 \times 10^2$ ks for NGC 1270 and $4.87 \times 10^2$ ks for NGC 1272.

\subsection{Spectral analysis}
Spectral analysis was conducted on NGC 1270 and NGC 1272 in the Perseus cluster where we extracted spectra of these objects using circles of radii $11\arcsec$. Local background spectra were obtained using annuli around the source and reaching out to $24\arcsec$. Photons with energies lower than $0.5$ keV or higher than $7.0$ keV were excluded due to low effective energy bands for {\it Chandra}.

\begin{figure*}
\begin{minipage}{0.49\hsize}
	\includegraphics[width=\columnwidth]{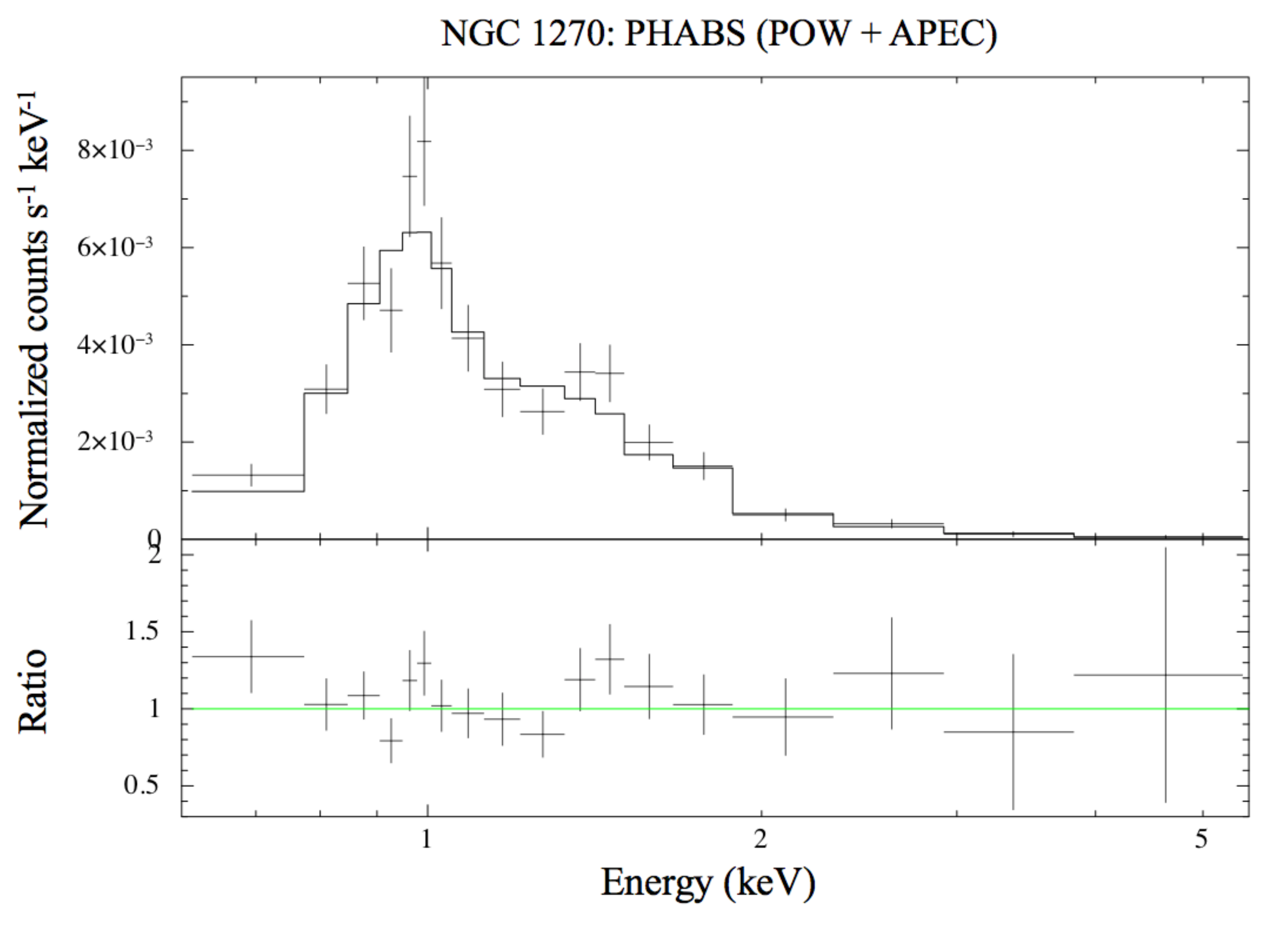}
\end{minipage}
\begin{minipage}{0.49\hsize}
	\includegraphics[width=\columnwidth]{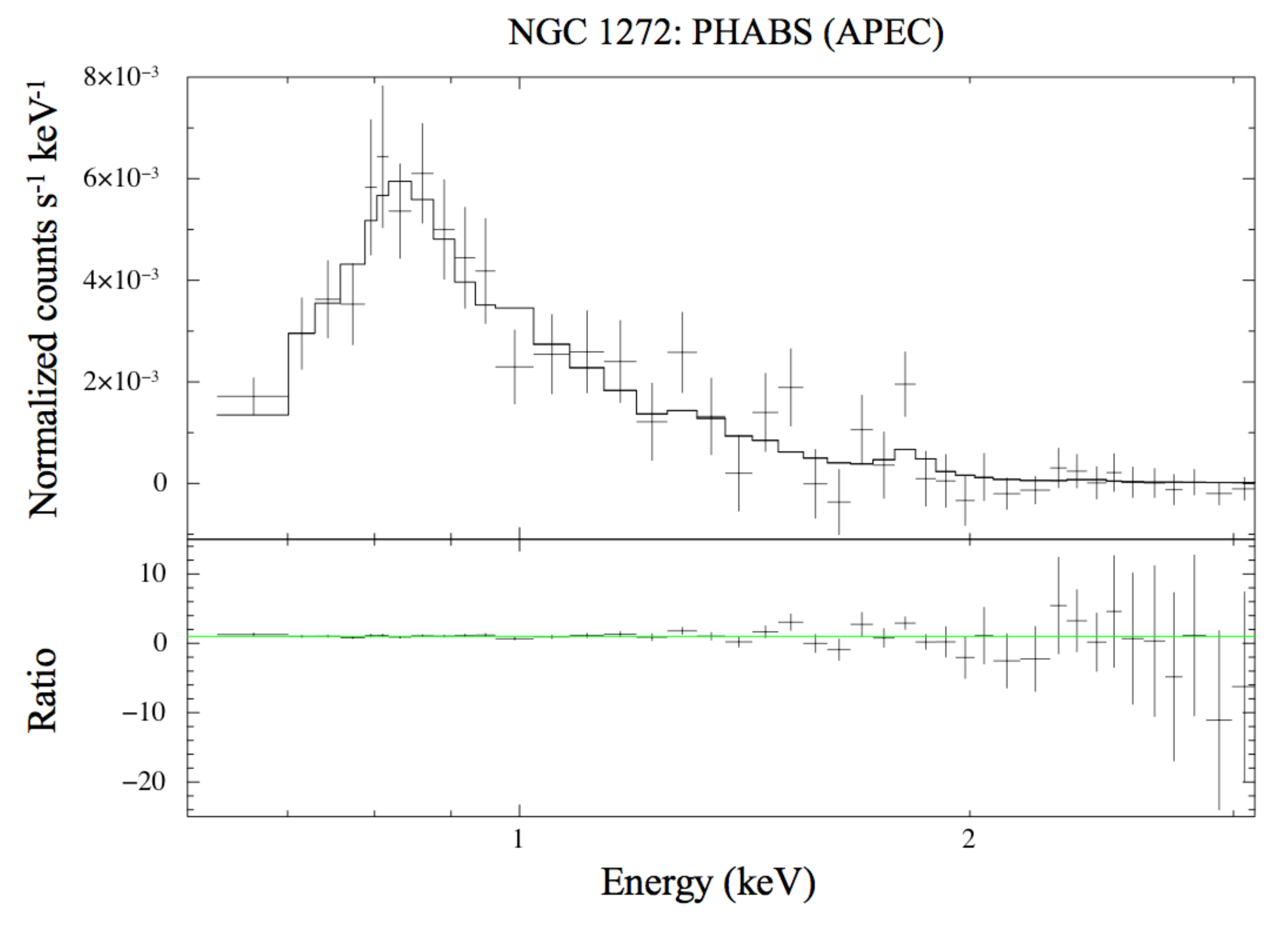}
\end{minipage}
\caption{X-ray spectral fitting. ({\it Left Panel}) X-ray spectrum of NGC 1270 in the Perseus cluster fitted with a galactic absorption applied to a power-law continuum and an {\sc apec} thermal model. The addition of a power-law component slightly improved the fit. ({\it Right Panel}) X-ray spectrum of NGC 1272 fitted with a galactic absorption applied to an {\sc apec}  thermal model.}
\label{fig:spec-fit}
\end{figure*}

Spectral models were fitted with {\sc xspec} version 12.10.0c. The local background spectra were subtracted from each region. The spectra of both galaxies were fitted with two emission models: an absorbed thermal component model ({\sc phabs} $\times$ {\sc apec}) and the combination of an absorbed power-law model and an absorbed thermal component ({\sc phabs $\times$ (pow + apec)}). The value of the absorbing column density of neutral hydrogen $N_{\mathrm{H}}$ was fixed to the LAB survey values \citep{2005AampA...440..775K} as $1.38\times 10^{21}$ cm$^{-2}$ which is consistent with galactic absorption in the directions. Besides, the redshifts used for these two objects were $0.0179$ as described before. Since in many cases, the value of photon index $\Gamma$ was fixed to $2.0$ \citep{1994MNRAS.268..405N} for typical active galactic nuclei (AGNs), this is used in our spectral fitting as well. The results of the spectral fitting minimizing the $\chi^2$ statistics for NGC 1270 and NGC 1272 are shown in Table~\ref{spec_table}. This shows that the best emission models are the latter one ({\sc phabs $\times$ (pow + apec)}) for NGC 1270 and the former one ({\sc phabs} $\times$ {\sc apec}) for NGC 1272 respectively. These spectra fitted with the best emission models are shown in Fig.~\ref{fig:spec-fit}.

 \begin{table*}
\centering
\caption{Spectral fit parameters for the two galaxies in the Perseus cluster. The data were fit between 0.5 and 7.0keV, minimizing the $\chi ^2$ statistics.}
\label{spec_table}
\begin{tabular}{llll}
\hline
Object & Parameter & Model ({\sc apec}) & Model ({\sc pow + apec}) \\ \hline
NGC 1270 & $N_{\mathrm{H}}$($10^{22}$cm$^{-2}$) & 0.138 fixed & 0.138 fixed \\
 & Photon index $\Gamma$ &  & 2.00 fixed \\
 & Power law normalization ($10^{-6}$) & & 2.21$^{+1.92}_{-1.94}$ \\
 & {\sc apec} normalization ($10^{-5}$) & 8.37$^{+1.30}_{-1.20}$& 6.43$^{+2.14}_{-2.13}$\\
 & kT (keV) & 1.00$_{-0.04}^{+0.04}$ & 0.99$_{-0.05}^{+0.04}$ \\
 & Z (Z$_\odot$) & 0.13$_{-0.03}^{+0.04}$  &  0.18$_{-0.06}^{+0.12}$\\
 & Count rate (Counts ks$^{-1}$) & 4.44 $\pm$ 0.21 & 4.44 $\pm$ 0.21 \\
 & $\chi^2/$ D.o.F. & $61.25/60=1.02$ & $59.96/59=1.02$ \\
 & $L_{0.5-2.0}$  ($10^{40}$ erg s$^{-1}$) & $3.61$ & $3.22$ ({\sc apec} component; unabsorbed) \\
 & $L_{0.5-2.0}$  ($10^{40}$ erg s$^{-1}$) &  & 0.86 (only {\sc pow}; $90\%$ upper limit; unabsorbed)\\
 & $L_{2-10}$ ($10^{40}$ erg s$^{-1}$) &  & 0.94 (only {\sc pow}; $90\%$ upper limit; unabsorbed) \\ \hline
Object & Parameter & Model ({\sc apec}) & Model ({\sc pow + apec}) \\ \hline
NGC 1272 & $N_{\mathrm{H}}$($10^{22}$cm$^{-2}$) & 0.138 fixed & 0.138 fixed \\
 & Photon index $\Gamma$ &  & 2.00 fixed \\
 & Power law normalization ($10^{-6}$) & & 0$^{+0.50}_{-0}$ \\
 & {\sc apec} normalization ($10^{-5}$) & 3.00$_{-1.89}^{+2.04}$ & 3.00$^{+2.04}_{-1.85}$ \\
 & kT (keV) & 0.63$_{-0.04}^{+0.04}$ & 0.63$_{-0.04}^{+0.04}$ \\
 & Z (Z$_\odot$) & 0.35$_{-0.16}^{+0.69}$ & 0.35$_{-0.16}^{+0.69}$ \\
 & Count rate (Counts ks$^{-1}$) & 2.68 $\pm$ 0.24 & 2.68 $\pm$ 0.24 \\
 & $\chi^2/$ D.o.F. & $227/227=1.00$ & $227/226=1.00$ \\
 & $L_{0.5-2.0}$ ($10^{40}$ erg s$^{-1}$) & 2.51 & 2.51 ({\sc apec} component; unabsorbed) \\
 & $L_{0.5-2.0}$  ($10^{40}$ erg s$^{-1}$) &  & 0.13 (only {\sc pow}; $90\%$ upper limit; unabsorbed)\\
 & $L_{2-10}$ ($10^{40}$ erg s$^{-1}$) &  & 0.14 (only {\sc pow}; $90\%$ upper limit; unabsorbed)
\end{tabular}
\end{table*}

\subsection{Surface Brightness}
\label{sec:surf}

\begin{figure}
 	\centering
\begin{minipage}{0.49\columnwidth}
	\includegraphics[width=\textwidth]{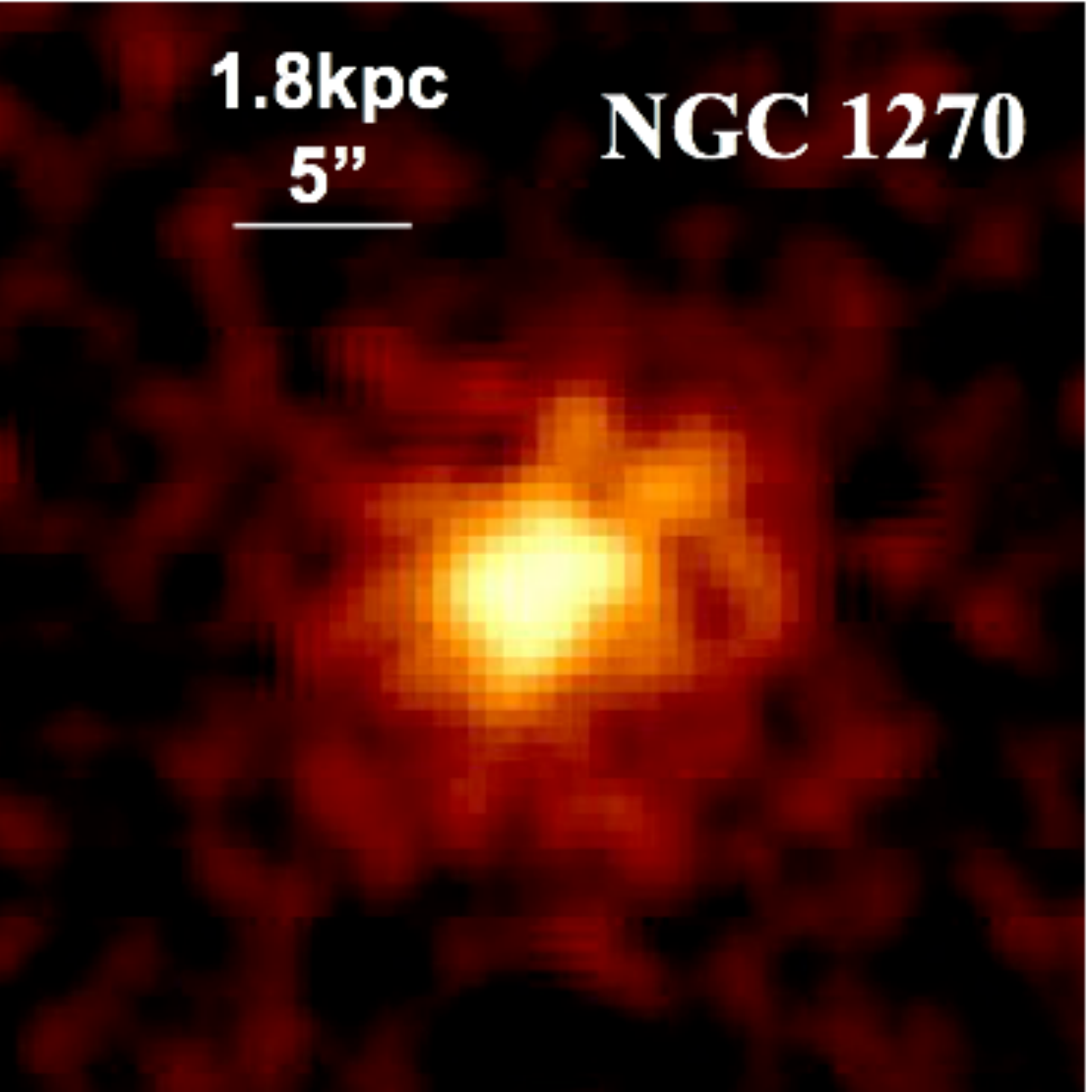}
\end{minipage}
\begin{minipage}{0.49\columnwidth}
	\includegraphics[width=\textwidth]{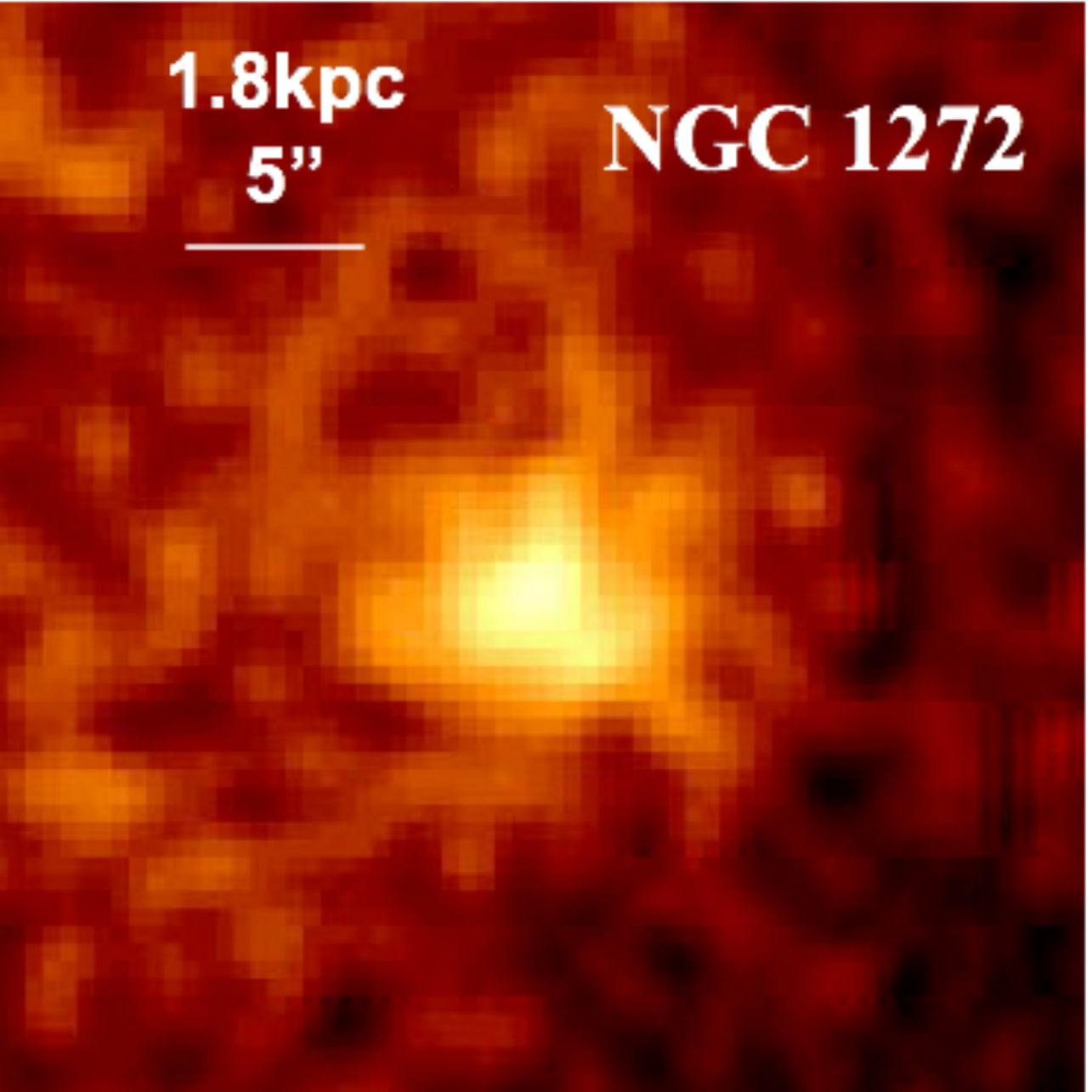}
\end{minipage}
\caption{Images from the ACIS-I detector of Chandra in $0.5 - 2.0$ keV, of NGC 1270 (left) and NGC 1272 (right). Note that the observations of these two galaxies are at high off-axis distances of $9.70\arcmin$ and $5.18\arcmin$ respectively.}
\label{fig:image_ObsPsf}
\end{figure}

In order to reveal whether these objects are consistent with being point-like AGN, or possessing a spatial extent, their point-spread functions (PSFs) were obtained and compared with the observed surface brightness. 
Since these two objects were located far off-axis at $9.70\arcmin$ ($\sim210$kpc) for NGC 1270 and $5.18\arcmin$ ($\sim110$kpc) for NGC 1272 from the aimpoint of Chandra ACIS-I observations, the PSF of the objects is asymmetric and broad. Therefore, we used point sources (i.e. distant AGNs) situated on the same off-axis circle as each galaxy on the chip as PSF. They were checked against SDSS (Sloan Digital Sky Survey) images to eliminate the possibility of spatially extended objects; member galaxies. We use the $0.5 - 2.0$ keV energy band for the surface brightness analysis; ACIS-I images are shown in Fig.~\ref{fig:image_ObsPsf}.

We assumed circular symmetry of the source in our analysis. As for NGC 1270, circular regions of radius $10\arcsec$ centred on the sources were used, divided into $10$ linearly spaced annuli to obtain the surface brightness profile. Background was extracted with a circle of radius $20\arcsec$ divided linearly into $20$ annuli to be eventually subtracted from the observed surface brightness profile. Regarding PSF, we assumed that within $5\arcsec$ around the center of the point-source object the surface brightness profile is well fitted with Gaussian function fixed at the center of the profile. Thus, we used a circular region of radius $5\arcsec$ divided linearly into $5$ annuli to extract the surface brightness profile of the point source. Background was extracted and subtracted in the same way as the observed NGC 1270 to finally produce its PSF profile.

NGC1272 is located at the cold front, and so extracting a background profile from circular annuli would cause the underestimation of the true background. Thus, we used fan-shaped annuli in ds9 analysis, toward the center of the Perseus cluster to the extent of the region with the uniform background mainly due to ICM so that more precise background information can be extracted. For NGC 1272, the surface brightness profile was extracted from a fan-shaped region of radius $15\arcsec$ and $15$ linear annuli while the background was obtained from the fan-shaped region of radius $20\arcsec$ and the $20$ annuli. The PSF of NGC 1272 was obtained similarly to that of NGC 1270.

\begin{figure*}
\begin{minipage}{0.49\hsize}
	\includegraphics[width=\columnwidth]{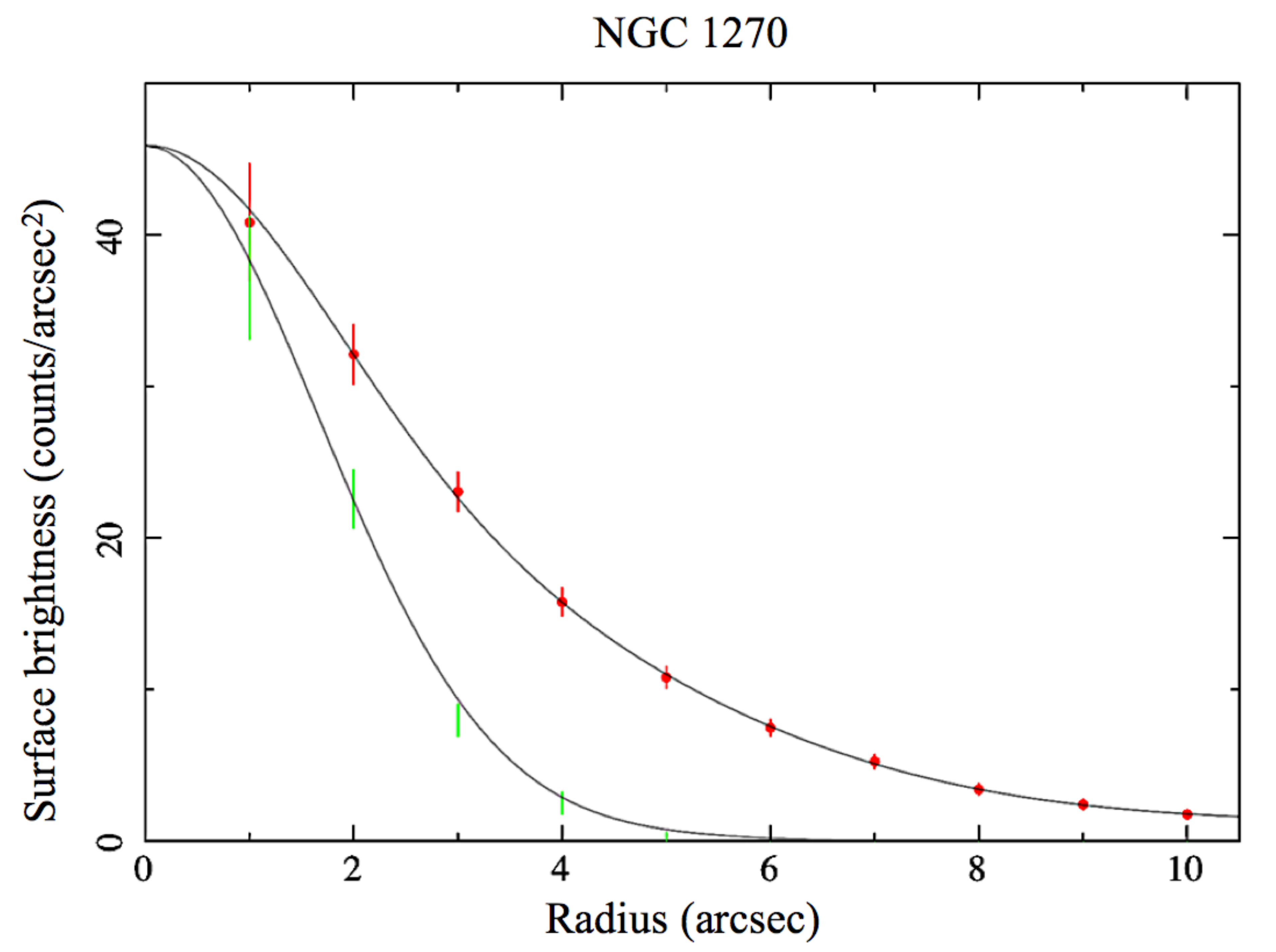}
\end{minipage}
\begin{minipage}{0.49\hsize}
	\includegraphics[width=\columnwidth]{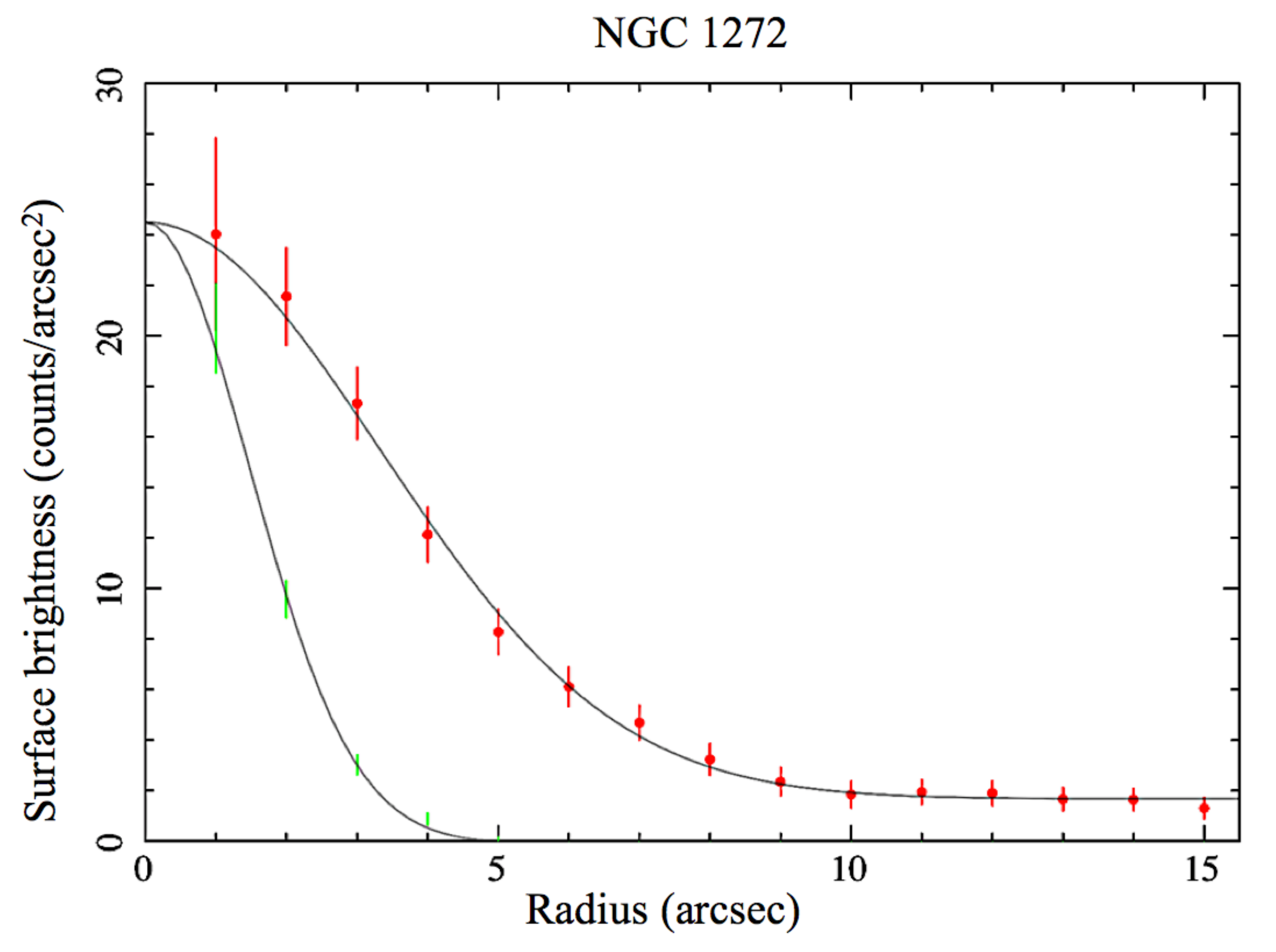}
\end{minipage}
\vspace*{+0.5cm} 
\caption{Background-subtracted X-ray surface brightness profiles and PSF profiles of the two galaxies in the $0.5$ to $2.0$ keV energy band. Left: NGC 1270 profile. Right: NGC 1272 profile. Surface brightness profiles are indicated with $1\,\sigma$ error bars. Profiles of the observation and the PSF for NGC 1272 and the PSF for NGC 1270 are fitted with a Gaussian function while the observation of NGC 1270 is fitted with a double-Gaussian function (see text). These fits are shown as solid lines. The observational profile data are plotted as red circles, while the PSFs are indicated with green error bars. The PSF profiles are normalized to match the observational profiles at the peak.}
 \label{fig:surf_qdp}
\end{figure*}

Hence, background-subtracted surface brightness profiles and PSFs, containing these data-points with $1\,\sigma$ error bars, were obtained for each source. The obtained background-subtracted X-ray surface brightness profiles as well as PSF profiles of both galaxies are presented in Fig.~\ref{fig:surf_qdp}. PSF profiles of the two were fitted with Gaussian function within $5\arcsec$ as mentioned above. While NGC 1272 was fitted well with a single-Gaussian function fixed at the center of the profile, NGC 1270 was better fitted with double-Gaussian functions, one of which is fixed with the given value of $\sigma$ of the Gaussian function already fitted onto its PSF. Then, $1\,\sigma$ value of each Gaussian function for both galaxies was extracted so that we can estimate the sizes of minicoronae, the spatially extended thermal components. (See Section~\ref{sec:corona-size}.)

\section{Results}
\label{sec:results}
\subsection{The presence of minicoronae}
As shown in Table~\ref{spec_table}, both galaxies demonstrated that they have highly significant thermal components in their spectra in comparison between unabsorbed thermal (minicorona) luminosity $L_{0.5-2.0}$ and unabsorbed power-law (point-source; AGN) $90\%$ upper-limit luminosity $L_{2-10}$. Especially for NGC 1272, given that the best spectral fit was with only an {\sc apec} thermal component model, NGC 1272 did not show any significant power-law component instead showed exceptionally strong dominance of thermal emission, indicating the ratio of soft thermal X-ray emission to the harder power-law emission as $L_{0.5-2.0}/L_{2-10}=17.3$. Note that the photon index of the power-law in the combination of power-law and thermal components was fixed to $\Gamma=2.0$ \citep{1994MNRAS.268..405N} in both galaxies so that the fitting program can constrain the results.

Moreover, both minicoronae showed the relatively cool temperature of $0.99$ keV for NGC 1270 and $0.63$ keV for NGC 1272, which were consistent with the past studies where the typical minicorona temperature is estimated as about $1$ keV \citep{2001ApJ...555L..87V, 	2005ApJ...619..169S, 	2002ApJ...578..833Y}.

\subsection{The radius of minicoronae}
\label{sec:corona-size}
As discussed in section \ref{sec:surf}, we assume an extent of $5\arcsec$ for the PSF surface brightness profiles is approximated by a Gaussian function. As Fig.~\ref{fig:surf_qdp} displays, as for NGC 1270, the double-Gaussian fitting of its surface brightness profile indicated one sigma value of the PSF component $\sigma_{\mathrm{PSF}} = 1.66\pm0.10\arcsec$, and normalization value $H$ was $H_{\mathrm{PSF}} = 20.1\pm 5.8$ while the other Gaussian function for the thermal corona component showed $\sigma_{\mathrm{c}}=3.63\pm 0.23\arcsec$ and $H_{\mathrm{c}} = 24.6 \pm 2.5$. Note that the flux ratio of AGN component to thermal component given by this analysis is consistent with the luminosity ratio in $0.5 - 2.0$ keV of the power-law to {\sc apec} component shown by spectral fitting in Table~\ref{spec_table}.

Likewise, as for NGC 1272, single-Gaussian fitting showed $\sigma_{\mathrm{PSF}}$ as $1.47\pm0.06\arcsec$ while the Gaussian fitting for observational surface brightness demonstrated $\sigma_{\mathrm{obs}}=3.32\pm0.15\arcsec$, providing the deconvolved $\sigma_{\mathrm{c}}$ value as $2.98\pm0.17\arcsec$.

Here we assume that the minicoronae are spherical, uniform and constant density gas where luminosity density is identical anywhere inside the coronal gas sphere. The projected surface brightness profile of this gas is expressed as
\begin{equation}
	I(R) = 2j_0 \sqrt{r_{\mathrm{c}}^2 - R^2},
\end{equation}
where $r_{\mathrm{c}}$, $j_0$ and $R$ are minicorona radius, constant luminosity density and the projected radius from the center of the corona respectively. The profile of the gas sphere is compared with the inner part of the Gaussian function of $\sigma_c$ for each minicorona (i.e. within $3\arcsec$ radius) to determine the (approximate) values of the minicorona radii $r_{\mathrm{c}}$ for both galaxies as shown in Table~\ref{results}.

 \begin{table*}
\centering
\caption{The physical properties of minicoronae in NGC 1270 and NGC 1272. The major uncertainties here are mainly due to systematics.}
\label{results}
\begin{tabular}{llllllllllll}
\hline
Object & $M_{\mathrm{BH}}$ & $kT$ &$r_{\mathrm{c}}$ & $n_{\mathrm{H,c}}$ &$n_{e,\mathrm{ICM}}$&$kT_{\mathrm{ICM}}$& $\lambda_{e,\mathrm{ICM}} $&$M_{\mathrm{c}}$ & $\dot M_{*}$ &$v_{\mathrm{r,gal}}$&$c_s$\\
&[M$_{\sun}$]& [keV] & [kpc]& [cm$^{-3}$]&[cm$^{-3}$]&[keV]&[kpc]&[M$_{\sun}$]&[M$_{\sun}$\,yr$^{-1}$]&[km s$^{-1}$]&[km s$^{-1}$]\\ \hline
NGC 1270 & $6.0\times 10^9$ & $0.99$ & $1.4$ & $1.0\times 10^{-1}$ &$3.7\times 10^{-3}$&$6.2$& $3.2$& $3.1\times10^7$ & $9.1\times 10^{-2}$& $3.0 \times 10^2$&$5.0\times10^2$\\
 NGC 1272 & $2.0\times 10^9$ & $0.63$ & $1.2$ & $9.0\times 10^{-2}$ &$9.2\times 10^{-3}$&$4.5$& $0.69$&$1.7\times10^7$ & $1.0\times 10^{-1}$& $1.5 \times 10^3$&$4.0\times10^2$\\
\end{tabular}
\\
\begin{tabular}{lllllllllll}
\hline
Object & $t_{\mathrm{rep}}$ & $t_{\mathrm{vs}}$ &$t_{\mathrm{rps}}$&$t_{\mathrm{ev}}$&$t_{\mathrm{cool}}$&$r_{\mathrm{a}}$& $\dot M_{\mathrm{Bondi}}$& $t_{\mathrm{b,acc}}$&$L_{2-10}$&$L_{\mathrm{b,bol}}$ \\
&[yr]& [yr] & [yr]&[yr]&[yr] & [kpc] &[M$_{\sun}$\,yr$^{-1}$] & [yr] &[erg s$^{-1}$]&[erg s$^{-1}$]\\ \hline
NGC 1270 & $3.4\times10^8$ & $4.4\times 10^{6}$& $4.1\times10^7$&$1.3\times10^6$ &$6.1\times 10^{7}$ & $0.20$ &$4.2\times 10^{-2}$&$7.3\times10^8$&$9.4\times10^{39}$ & $2.4\times10^{44}$ \\
 NGC 1272 & $1.7\times10^8$ &$4.1\times 10^{6}$& $4.3\times10^6$&$1.8\times10^6$ &$2.1\times10^7$ & $0.11$  &$8.1\times 10^{-3}$& $2.1\times10^9$& $1.4\times10^{39}$ & $4.6\times10^{43}$ \\
\end{tabular}
\end{table*}

\subsection{The central black hole mass}
\label{mass}
Using $M-\sigma$ relation, $M \propto \sigma^4$ \citep{2013ARAA..51..511K} and the values of central stellar radial velocity dispersions $\sigma_r$ of the two objects: $393$ km s$^{-1}$ for NGC 1270 \citep{2012Natur.491..729V} and $288$ km s$^{-1}$ for NGC 1272 (HyperLEDA), the central black hole masses $M_{\mathrm{BH}}$ of these sources were estimated as $6.0\times 10^9\ \mathrm{M}_{\sun}$ and $2.0\times 10^9\ \mathrm{M}_{\sun}$ respectively, as indicated in Table~\ref{results}. Both are supermassive black holes approaching that in NGC 1277 which has an ultramassive central black hole with a mass of $1.7 \times 10^{10}\ \mathrm{M}_{\sun}$ \citep{2013MNRAS.431L..38F}.

\subsection{The density and mass of minicoronae}
The normalization value $A$ of an {\sc apec} model is defined in CGS units as 
\begin{equation}
	A \equiv \frac{10^{-14}}{4\pi \left[D_A (1+z)\right]^2}\int n_e n_{\mathrm{H}} dV,
\end{equation}
where $D_A$, $z$, $n_e$ and $n_{\mathrm{H}}$ are the angular diameter distance, redshift, electron number density and proton number density respectively.

Assuming that the corona consists of a gas sphere of constant density, radius $r_{\mathrm{c}}$ and the normalization value shown in Table~\ref{fig:spec-fit}, the gas density of those minicoronae $n_{\mathrm{H,c}}$ were derived. Note that we assumed the ratio of $n_e$ to $n_{\mathrm{H}}$ is $1.2$. Then, the mass of the minicoronae $M_{\mathrm{c}}$ were also calculated. They are shown in Table~\ref{results}.

\subsection{The origin and replenishment of minicorona}
\label{replenishment}

The most simple way to describe the origin of the minicoronae is by replenishment due to the stellar mass loss as planetary nebula. \citet{2007ApJ...665.1038C} showed the stellar mass loss rate induced by stellar evolution in elliptical galaxies was estimated as
\begin{equation}
	\dot M _*(t) \sim 1.5 \times \frac{L_{\mathrm{B}}}{10^{11} L_{\mathrm{B}\sun}}t_{15}^{-1.3}\ \mathrm{M}_{\sun}\,\mathrm{yr}^{-1},
\end{equation}
where $L_{\mathrm{B}}$, $L_{\mathrm{B}\sun}$ and $t_{15}$ are the present galaxy blue luminosity and that of sun and the age of the object in $15$ Gyr units.

Here we estimated the age of these two galaxies at about $13.5$ Gyr. The term $L_{\mathrm{B}}/L_{\mathrm{B}\sun}$ was estimated as $1.6 \times 10^{10}$ for NGC 1270 and $5.6 \times 10^{10}$ for NGC 1272. Since we focus on studying minicorona, the central compact gaseous region in the elliptical galaxy, we derived galaxy blue luminosity within minicorona size so that we can estimate more accurate replenishment rate of gas for minicorona. For the given coronal radii, we estimated the fractions of blue luminosities in the minicoronae to those in the whole galaxies are $0.25$ for NGC 1270 and $0.10$ for NGC 1272 through the flux volume integral. Thus, we calculate the stellar mass loss rate $\dot M_*$ in the cores of both galaxies as well as replenishment time $t_{\mathrm{rep}}$ for the given $M_{\mathrm{c}}$ to be accumulated due to stellar mass loss. The replenishment time is calculated as $3.4\times 10^8$ yr and $1.7\times 10^8$ yr respectively. The full properties are shown in Table~\ref{results}.

\subsection{The depletion of minicoronal gas due to viscous stripping and evaporation by surrounding ICM}
\label{stripping}
There are two main transport processes that could be considered for depletion of coronal gas at the boundary between minicorona and ICM: viscous stripping and evaporation (thermal conductivity). First, we focus on the former process on minicorona.
\citet{1982MNRAS.198.1007N} found a viscous stripping time scale $t_{\mathrm{vs}}$ as follows:
\begin{align}
		t_{\mathrm{vs}}=\frac{4}{3} \left(\frac{r_{\mathrm{c}}}{v_{\mathrm{gal}}}\right) \left(\frac{\rho_{\mathrm{c}}}{\rho_{\mathrm{ICM}}}\right)\left(\frac{12}{R_e}+1\right)^{-1},
\end{align}
where $v_{\mathrm{gal}}$ is the speed of the galaxy relative to ICM, $\rho_{\mathrm{c}}$ and $\rho_{\mathrm{ICM}}$ are density of minicorona and ICM at the boundary between minicorona and surrounding ICM, and $R_e$ is the Reynolds number which is expressed as
\begin{equation}
	R_e = \frac{\rho_{\mathrm{ICM}} v_{\mathrm{gal}} r_{\mathrm{c}}}{\eta},
\end{equation}
where $\eta$ is dynamic viscosity of ICM which can be indicated as
\begin{equation}
\eta = 1410 \left(\frac{kT_{\mathrm{ICM}}}{5.0\ \mathrm{keV}}\right)^{\frac{5}{2}}\left(\frac{\ln \Lambda}{40}\right)^{-1}\mathrm{g\,cm}^{-1}\,\mathrm{s}^{-1},
\end{equation}
where $T_{\mathrm{ICM}}$ is the projected ICM electron temperature at the boundary between minicorona  surface and surrounding ICM, and $\ln \Lambda$ is Coulomb logarithm of ICM that can be expressed as
\begin{equation}
	\ln \Lambda = 37.3 + \ln \left[\left(\frac{kT_{\mathrm{ICM}}}{5.0\ \mathrm{keV}}\right)\left(\frac{n_{e,\mathrm{ICM}}}{10^{-3}\ \mathrm{cm}^{-3}}\right)^{-\frac{1}{2}}\right],
\end{equation}
where  $n_{e,\mathrm{ICM}}$ is the projected electron number density of ICM at that boundary. Note that those formulae of the parameters demonstrated above did not take the magnetic field into account. As noted in \citep{1962pfig.book.....S}, the mean free path of the electron in the ICM $\lambda_{e,\mathrm{ICM}}$ at that boundary can be estimated as
\begin{align}
\lambda_{e,\mathrm{ICM}}&=\frac{3^{\frac{3}{2}}(kT_{\mathrm{ICM}})^2}{4\pi^{\frac{1}{2}}e^4 n_{e,\mathrm{ICM}}\ln \Lambda}\\
		&\sim7.7\left(\frac{kT_{\mathrm{ICM}}}{5.0\ \mathrm{keV}}\right)^2 \left(\frac{n_{e,\mathrm{ICM}}}{10^{-3}\ \mathrm{cm}^{-3}}\right)^{-1},
\end{align}
where $e$ is the elementary electric charge.

We used Equation $4$ and $5$ in \cite{2003ApJ...590..225C} to derive these projected ICM properties of the temperature $kT_{\mathrm{ICM}}$, electron number density $n_{e,\mathrm{ICM}}$ and the mean free path $\lambda_{e,\mathrm{ICM}}$. These properties are shown in Table~\ref{results}.

Considering the size of the Perseus cluster, $\sim1$ Mpc, which provides the limitation of the dispersion of the radial distance between each Perseus member galaxy and us, it can be estimated that the recession velocity of each galaxy and the cluster peculiar velocity are almost identically common among the Perseus member galaxies. Thus, assuming that central ICM and central galaxy NGC 1275 of the Perseus cluster are static against NGC 1270 and NGC 1272, galaxy radial peculiar velocity relative to surrounding ICM $v_{\mathrm{r,gal}}$ can be derived from the difference of observed redshift between the object and central NGC 1275. Therefore, the galaxy radial velocities $v_{\mathrm{r,gal}}$ were calculated as $300$ km s$^{-1}$ for NGC 1270 and $1450$ km s$^{-1}$ for NGC 1272 where both galaxies are moving towards us.

Since $v_{\mathrm{gal}}$ indicated above needs to be three-dimensional velocity of the galaxy, here the given radial velocities were substituted being used as the lower limit in the estimation of viscous stripping time scale. Hence, viscous stripping time can be expressed as the function of the Reynolds number:
\begin{align}
		t_{\mathrm{vs}}&\sim 1.6\times 10^8 \left(\frac{r_{\mathrm{c}}}{1.0\ \mathrm{kpc}}\right)\left(\frac{v_{\mathrm{gal}}}{10^3\ \mathrm{km\ s}^{-1}} \right)^{-1}\nonumber\\
			&\ \ \ \ \ \times\left(\frac{n_{\mathrm{H,c}}}{0.1\ \mathrm{cm}^{-3}}\right)\left(\frac{n_{e,\mathrm{ICM}}}{10^{-3}\ \mathrm{cm}^{-3}} \right)^{-1}\left(\frac{12}{R_e}+1\right)^{-1}\mathrm{yr}\\
		&\sim \begin{cases}
			2.0\times 10^8 \left(\frac{12}{R_e}+1\right)^{-1}\mathrm{yr} & (\text{NGC 1270})\\
			1.3\times 10^7 \left(\frac{12}{R_e}+1\right)^{-1}\mathrm{yr} & (\text{NGC 1272}).
		\end{cases} 	
\end{align}

Then, if the magnetic field is neglected, those parameters can be calculated as follows: $\ln \Lambda \sim 36.8$ and $36.0$, $\eta \sim 2620$ and $1210$, and $R_e > 0.26$ and $5.8$ for NGC 1270 and NGC 1272 respectively. As shown in Table~\ref{results}, it finally gave the value of viscous stripping time as 
\begin{equation}
		t_{\mathrm{vs}}\sim \begin{cases}
			4.4\times 10^6\ \mathrm{yr}  & (\text{NGC 1270})\\
			4.1\times 10^6\ \mathrm{yr}  & (\text{NGC 1272}).
		\end{cases} 
\end{equation}

Thus, it showed that compared to the given replenishment time scale due to stellar mass loss discussed in Section~\ref{replenishment}, the depletion time scale via viscous stripping demonstrated a shorter value by a factor of $\sim 100$ in the case without magnetic field. This result means that stripping must be strongly suppressed by a factor of $\sim 100$ so that minicorona can be observed.

We should note that the viscous stripping is more crucial than ram pressure stripping for the depletion of minicoronal gas. As discussed in \citep{1986RvMP...58....1S}, the ram pressure stripping time scale $t_{\mathrm{rps}}$ can be approximated as
\begin{align}
	t_{\mathrm{rps}}&\sim \frac{r_{\mathrm{c}}}{v_{\mathrm{gal}}}\left(\frac{2\rho_{\mathrm{c}}}{\rho_{\mathrm{ICM}}}\right)^{\frac{1}{2}}\\
			&\sim 1.6\times10^7 \left(\frac{n_{\mathrm{H,c}}}{0.1\ \mathrm{cm}^{-3}}\right)^{\frac{1}{2}}\left(\frac{n_{e,\mathrm{ICM}}}{10^{-3}\ \mathrm{cm}^{-3}} \right)^{-\frac{1}{2}}\nonumber\\
			&\ \ \ \ \ \ \ \ \ \ \ \ \ \ \ \ \ \ \ \times\left(\frac{v_{\mathrm{gal}}}{10^3\ \mathrm{km\ s}^{-1}} \right)^{-1}\left(\frac{r_{\mathrm{c}}}{1.0\ \mathrm{kpc}}\right)\ \mathrm{yr}.
\end{align}
If the magnetic field is neglected, this is calculated as $< 4.1\times10^7$ yr and $< 4.3\times10^6$ yr respectively, showing that the time scale for ram pressure stripping is longer than that of viscous stripping.

Now we focus on the second main transport process, evaporation. As noted in \citep{1977Natur.266..501C}, the evaporation rate $\dot M_{\mathrm{ev}}$ of the minicoronal gas can be estimated as
\begin{align}
	\dot M_{\mathrm{ev}} &\sim \frac{16\pi\mu m_{\mathrm{p}} \kappa r_{\mathrm{c}}}{25k}\\
	&\sim 9.0 \left(\frac{kT_{\mathrm{ICM}}}{5.0\ \mathrm{keV}}\right)^{\frac{5}{2}}\left(\frac{r_{\mathrm{c}}}{1.0\ \mathrm{kpc}}\right)\left(\frac{\ln \Lambda}{40}\right)^{-1}\msunyreq,
\end{align}
where $\mu$ and $\kappa$ are the mean molecular weight and the thermal conductivity for a hydrogen plasma respectively. Thus, the evaporation time scale $t_{\mathrm{ev}}$ of the minicoronal gas is given by
\begin{equation}
	t_{\mathrm{ev}}=\frac{M_{\mathrm{c}}}{\dot M_{\mathrm{ev}}}.
\end{equation}
As shown in Table~\ref{results}, if the magnetic field is neglected, the evaporation time is calculated as $1.3\times10^6$ yr and $1.8\times10^6$ yr respectively, indicating the relation: $t_{\mathrm{ev}}<t_{\mathrm{vs}}<t_{\mathrm{rps}}$.

\subsection{The radiative cooling of minicoronal gas and heating by Type Ia supernova}
\label{cooling}
Assuming that the temperature of protons is similar to that of electrons, the radiative cooling time $t_{\mathrm{cool}}$ of each minicorona can be estimated as
\begin{align}
	t_{\mathrm{cool}} &\sim \frac{3N_p kT}{2L_{\mathrm{bol}}}\\
		&\sim 2.0 \times 10^7 \left(\frac{M_{\mathrm{c}}}{10^7\ \mathrm{M}_{\sun}}\right)\nonumber\\
		&\ \ \ \ \ \times\left(\frac{kT}{1.0\ \mathrm{keV}}\right)\left(\frac{L_{\mathrm{bol}}}{10^{41}\ \mathrm{erg}\ \mathrm{s}^{-1}} \right)^{-1}\mathrm{yr},
\end{align}
where $N_p$ is the number of particles in the minicorona, and $L_{\mathrm{bol}}$ is bolometric luminosity of the objects. As shown in Table~\ref{results}, the cooling time for each galaxy is calculated as $6.1\times 10^7$ yr and $2.1\times 10^7$ yr respectively.

Using Equation 12 in \citet{2007ApJ...665.1038C}, the heating rate by Type Ia supernova $L_{\mathrm{SN}}$ can be estimated as $L_{\mathrm{SN}} < 3.0 \times 10^{40}\ \mathrm{erg\ s}^{-1}$ for NGC 1270 and $L_{\mathrm{SN}} < 3.4 \times 10^{40}\ \mathrm{erg\ s}^{-1}$ for NGC 1272 respectively. Note that they are upper limits, and the fraction of the total energy released by a supernova event being used for heating minicoronal gas should be less than $1$ in reality. Comparing these heating rates with the bolometric luminosity of the objects, it is seen that supernovae are not sufficient enough as a heat source to offset cooling alone. Heating by AGN and a balance between cooling and heating are discussed in Section~\ref{sec:discussion}.

\subsection{Comparison between the observed power-law luminosity and Bondi Accretion Luminosity}
\label{bondi}

Using the derived black hole masses, the unabsorbed bolometric luminosity of these two objects were estimated, assuming Bondi accretion so that an evaluation of their AGN activities can be obtained. The unabsorbed bolometric luminosity $L_{\mathrm{b,bol}}$ was calculated as follows:
\begin{equation}
L_{\mathrm{b,bol}}=\epsilon \dot M_\mathrm{Bondi} c^2.
\end{equation}
The Bondi accretion rate $\dot M_{\mathrm{Bondi}}$ is given by
\begin{equation}
\dot M_{\mathrm{Bondi}} = 4\pi \lambda_B \rho (r_{\mathrm{a}})\frac{G^2 M^2_{\mathrm{BH}}}{c_s^3},
\end{equation}
where the medium sound speed $c_s$ was estimated as
\begin{equation}
	c_s=\sqrt{\frac{\gamma kT}{\mu m_p}},
\end{equation}
and $\epsilon$, $\lambda_B$, $\gamma$, $m_p$, $G$, $c$ and $\rho$ are radiative efficiency, adiabatic correction, adiabatic index, proton mass, gravitational constant, speed of light and ambient density at the accretion radius $r_{\mathrm{a}}$ respectively \citep{1952MNRAS.112..195B}. Here $\mu$ was assumed to be 0.62. For this analysis, it was assumed that the accretion process is adiabatic while the minicorona gas distribution is isothermal. As shown in Table~\ref{spec_table}, the values of $kT$ were derived from our results of spectral fitting: $0.99$ keV for NGC 1270, and $0.63$ keV for NGC 1272 respectively. In addition, $\gamma=5/3$ for an adiabatic process which provide the value of $\lambda_B$ as $0.25$. The radiative efficiency $\epsilon$ was assumed to be $0.1$.

The ambient density $\rho$ is expressed as
\begin{equation}
\rho(r_{\mathrm{a}})=m_p n_{\mathrm{H}}(r_{\mathrm{a}}),
\end{equation}
where accretion radius $r_{\mathrm{a}}$ is given by
\begin{align}
r_{\mathrm{a}} &=\frac{2GM_{\mathrm{BH}}}{c_s^2}\\
&\sim 33 \left(\frac{M_{\mathrm{BH}}}{10^9\ \mathrm{M}_{\sun}} \right) \left(\frac{kT}{1.0\ \mathrm{keV}} \right)^{-1}\mathrm{pc},
\end{align}
where $r_{\mathrm{a}}$ can be estimated as $0.20$ kpc and $0.11$ kpc for each galaxy respectively. Since in our paper, it is assumed that the minicorona is comprised of constant density gas sphere, given that $r_{\mathrm{a}} < r_{\mathrm{c}}$, the derived minicorona proton number density $n_{\mathrm{H,c}}$ can be directly used as $n_{\mathrm{H}}(r_{\mathrm{a}})$.

Therefore, the Bondi accretion bolometric luminosity can be estimated as
\begin{align}
	L_{\mathrm{b,bol}}&\sim 6.4\times10^{42} \left(\frac{n_{\mathrm{H,c}}}{0.1\ \mathrm{cm}^{-3}} \right)\nonumber\\
	&\ \ \ \ \ \times\left(\frac{M_{\mathrm{BH}}}{10^9\ \mathrm{M}_{\sun}} \right)^2\left(\frac{kT}{1.0\ \mathrm{keV}} \right)^{-\frac{3}{2}}\mathrm{erg\ s}^{-1}.
\end{align}
Thus, those bolometric luminosities for both galaxies were calculated as well as the medium sound speed and the Bondi accretion rate. The Bondi accretion time scale $t_{\mathrm{b,acc}}$ of the minicoronae was estimated by using
\begin{equation}
	t_{\mathrm{b,acc}}=\frac{M_\mathrm{c}}{\dot M_\mathrm{Bondi}}.
\end{equation}
As shown in Table~\ref{results}, the Bondi accretion bolometric luminosity is estimated as $2.4\times 10^{44}$ erg s$^{-1}$ and $4.6\times 10^{43}$ erg s$^{-1}$, and the Bondi accretion time scale is calculated as $7.3\times 10^8$ yr and $2.1\times 10^9$ yr respectively.

Compared with the power-law $90\%$ upper limit luminosities $L_{2-10}$ shown in Table~\ref{spec_table}, even if considering the hard X-ray bolometric corrections, $\sim10\%$ \citep{2009MNRAS.392.1124V}, the Bondi accretion luminosity was larger by a factor of $\sim10^3$ than the observed power-law luminosity in both galaxies. This result clearly shows that these objects are significantly under-luminous in X-rays. 

\section{Discussion and conclusion}
\label{sec:discussion}
To begin with, note that \citet{2007MNRAS.382..895S} showed that LMXB contributions to the X-ray emissions are likely to be small under the observing conditions of the Perseus core galaxies. Thus, X-ray emissions are dominated by AGN activity and nuclear gas without any significant LMXB emission. Besides, it should also be noted that in the comparison between Bondi accretion powered bolometric luminosity and observed power-law upper limit luminosity, the latter luminosity was given in the energy range $2.0-10.0$ keV. Thus, for more precise comparison, harder X-ray observational data could be also taken into account though it would have only a negligible effect.

As shown in Section~\ref{replenishment} and \ref{stripping}, our study showed that, neglecting magnetic field, the depletion time scale via viscous stripping was shorter by a factor of $\sim 100$ than the replenishment time scale due to stellar mass loss. It also demonstrates the viscous stripping process must be strongly suppressed by a factor of $\sim 100$ so that minicorona can be long-lived. Our result is consistent with the past study of S07, and magnetic fields could provide a plausible explanation for the required strong suppression.

The other transport process discussed, evaporation, can work for galaxies moving slowly in the ICM. We showed that evaporation was $\sim3$ times faster than viscous stripping for minicoronae in massive early-type galaxies such as NGC 1270 and NGC 1272, demonstrating consistency with the study of \citet{1982MNRAS.198.1007N}. Consequently, it is shown that the relation $t_{\mathrm{ev}}<t_{\mathrm{vs}}<t_{\mathrm{rps}}$ can be applied for minicoronae, if the magnetic field is neglected. That is why heat conduction needs to be suppressed by a factor of $> 100$, similar to the study of S07. This conclusion also reinforced the argument that X-ray minicoronae should be magnetized so that all these depletion processes can be significantly suppressed.

We show the estimated mean free path of electrons in the medium around the galaxies is comparable to the size of minicoronae, if the magnetic field is neglected. This indicates that minicoronal gas is collisionless. Collisions would be suppressed by a magnetic field. The radiative cooling time scales were shown to be shorter than the age of these galaxies by a factor of $>100$. This implies that there should be a heating source in the galactic core which can offset such rapid cooling. As shown in Section~\ref{cooling}, Type Ia supernova cannot provide sufficient heating for minicorona alone to completely offset the radiative cooling. Instead, an AGN is an obvious heating source.

Now we discuss the reason why these objects are significantly under-luminous in X-rays by a factor of $10^3$ as shown in Section~\ref{bondi}. There are two scenarios to address this problem: (1) Accretion is hindered, so accretion rate appears to be low, or (2) Accretion actually occurs while being hindered, but radiation production is inefficient.

The former case could be induced by ICM ram pressure. However, this scenario would not work well for these galaxies since the soft X-ray emitting gas is present in the core of both galaxies. This implies that the ram pressure due to the motion of galaxies relative to surrounding ICM is not strong enough to push the gas away. Furthermore, the bent jet (WAT; wide-angle tale radio source) observed in NGC 1272 shows that the gas is not much disturbed within the core of $\sim2$ kpc while traversing the minicorona \citep{2014MNRAS.442..838M}. As also discussed in Section~\ref{stripping}, NGC 1272 is estimated to move at least $1450$ km s$^{-1}$ relative to the surrounding ICM, which causes the strong ram pressure and hence the WAT outside the central region of $2$ kpc. Thus, this scenario does not apply in its small minicorona of radius $1.2$ kpc.

On the other hand, it is well known that low luminosity AGNs have radiatively inefficient accretion flows (RIAFs). \citet{1995ApJ...452..710N} predicted that the luminosity is approximately proportional to the square of the accretion rate, and therefore the radiative efficiency drops as the accretion rate is reduced. Hence, once the accretion rate $\dot M$ drops, the luminosity will be significantly decreased by $\dot M^2$ rather than linear drop. As suggested by \citet{1995MNRAS.277L..55F} this could be a key to explaining the huge luminosity difference of 3 orders of magnitude. Moreover, since NGC 1272 has the radio jet, when the total energy of the source goes up, the fraction of energy in the kinetic jet could show much larger increase. Therefore, it might be possible that accretion rate may be low and at the same time radiatively inefficient.

There are no good examples where the Bondi accretion rate has been confirmed by observation \citep[see e.g.][]{2018MNRAS.477.3583R,		2014ARAA..52..529Y, 	2014ApJ...780....9W}. We should note that the calculated Bondi accretion rate and the bolometric Bondi accretion luminosity in TabIe~\ref{results} are lower limits since we assumed the constant density. Jets and winds from the center may heat and choke the inflow and together with radiatively inefficient accretion make any accretion luminosity low or even unobservable. If either jet or wind appears in $50\%$ of the time and provides a power of $10^{41}$ erg s$^{-1}$, it will offset the luminosity from minicoronae and so maintain a balance between heating and cooling. We note again that NGC 1272 has the WAT which bears similarity to the situation in one (NGC 4874) of the two minicoronae in the Coma cluster \citep{2014MNRAS.439.1182S}. We conclude that minicoronae need to make both a heating - cooling balance as well as a replenishment - stripping one. Balancing both enables them to maintain the relatively cool and dense gas in the middle of the harsh ICM environment.

\section*{Acknowledgements}
We thank H. Russell for helpful discussions and the referee for comments. NA acknowledges the Gates Cambridge Scholarship from the Bill \& Melinda Gates Foundation and the support from Honjo International Scholarship Foundation. ACF acknowledges ERC Advanced Grant 340442. SAW was supported by an appointment to the NASA Postdoctoral Program at the Goddard Space Flight Center, administered by the Universities Space Research Association through a contract with NASA.



\bibliographystyle{mnras}
\bibliography{Minicorona} 






\bsp	
\label{lastpage}
\end{document}